\documentclass[lettersize,journal]{IEEEtran}
\usepackage[utf8]{inputenc}
\usepackage{booktabs}
\usepackage{tabularx}
\usepackage{amsmath,amsfonts}
\usepackage{graphicx}
\usepackage{algorithm}
\usepackage{algpseudocode}
\usepackage{array}
\usepackage[caption=false,font=normalsize,labelfont=sf,textfont=sf]{subfig}
\usepackage{textcomp}
\usepackage{stfloats}
\usepackage{url}
\usepackage{verbatim}
\usepackage{cite}
\usepackage{float}
\begin{document}

\title{6Diffusion: IPv6 Target Generation Using a Diffusion Model with Global-Local Attention Mechanisms for Internet-wide IPv6 Scanning}

\author{Nabo He, \and DanDan Li, \and Xiaohong Huang
\thanks{This work was supported by National Natural Science Foundation of China under Grant No.62471070. (Corresponding Author: Dandan Li)}
\thanks{Nabo He and Dandan Li are with the School of Computer Science (National Pilot Software Engineering School), Beijing University of Posts and Telecommunications, Beijing 100876, China. Xiaohong Huang is State Key Laboratory of Networking and Switching Technology, Beijing University of Posts and  Telecommunications,  Beijing 100876, China. (e-mail: henabo@bupt.edu.cn; dandl@bupt.edu.cn; huangxh@bupt.edu.cn).}
}

\maketitle

\begin{abstract}
Due to the vast address space of IPv6, the brute-force scanning methods originally applicable to IPv4 are no longer suitable for proactive scanning of IPv6. The recently proposed target generation algorithms have a low hit rate for existing IPv6 target generation algorithms, primarily because they do not accurately fit the distribution patterns of active IPv6 addresses. This paper introduces a diffusion model-based IPv6 target generation algorithm called 6Diffusion. 6Diffusion first maps addresses to vector space for language modeling, adds noise to active IPv6 addresses in the forward process, diffusing them throughout the entire IPv6 address space, and then performs a reverse process to gradually denoise and recover to active IPv6 addresses. We use the DDIM sampler to increase the speed of generating candidate sets. At the same time, we introduce the GLF-MSA (Global-Local Fusion Multi-Head Self-Attention) mechanism to adapt to the top-down global allocation pattern of IPv6 addresses and the local characteristics of IPv6 address segments, thus better learning the deep-level features of active IPv6 addresses. Experimental results show that compared to existing methods, 6Diffusion can generate higher quality candidate sets and outperforms state-of-the-art target generation algorithms across multiple metrics.
\end{abstract}


\section{Introduction}
\IEEEPARstart{I}{P} address scanning is an important means of network research. It has applications in network service measurement \cite{hu2018measuring, durumeric2015search, czyz2016don, durumeric2014internetwide}, network topology discovery \cite{beverly2016yarrping}, cybersecurity, digital asset vulnerability analysis \cite{dawood2012ipv6}, address analysis \cite{foremski2016entropy} and so on. IP address scanning has made significant advancements in IPv4, however, due to the imminent exhaustion of IPv4 addresses, more institutions are turning towards the next-generation Internet Protocol \cite{w3techs2024}, IPv6, which offers an exponentially larger address space. Yet, the brute-force scanning methods originally used for IPv4 are not applicable to the vast address space of IPv6 \cite{gasser2018clusters}, making the scanning of the IPv6 address space an urgent current requirement.

To resolve the challenges in IPv6 address scanning, researchers have proposed IPv6 target generation algorithms (TGAs). It is customary to gather a set of known active IPv6 addresses, which are commonly known as the seed sets. Algorithms are then employed to study their salient features and generate a higher quality set of candidate addresses for subsequent scanning; thus, the design of the algorithm is of crucial.

However, this approach may present several issues: 1) Low hit rate, due to the vast IPv6 address space and the complexity of IPv6 addressing patterns, although RFC7707 \cite{chown2016network} has proposed six recommended patterns for IPv6 allocation, servers and routers are often allocated based on administrators' habits, and network administrators may have their own allocation methods, thus making the actual network allocation more complex. This leads to a lower hit rate for candidate sets. 2) Monolithic address patterns. Current IPv6 target generation algorithms largely utilize the DHC algorithm \cite{murdock2017target}, thus focusing mainly on the seed sets prefixes (which can easily become trapped in a region under dynamically adjusted algorithms, continuously generating addresses). For areas with scarce seed addresses, the generation effect is less satisfactory, which is also a challenge that IPv6 target address generation algorithms need to overcome.

To address these challenges, we introduce 6Diffusion, an IPv6 target generation algorithm that leverages the Diffusion Model\cite{ho2020denoising} to enhance the quality of candidate address sets. Our approach begins with semantic modeling of the IPv6 address space, translating discrete addresses into a continuous vector space. This allows for a more nuanced understanding of the semantic relationships between addresses. By employing a Diffusion Model, we can generate active IPv6 addresses that are not merely extensions of the seed set but are informed by a broader understanding of the address space's semantics.

6Diffusion's utilizing the Transformer\cite{vaswani2017attention} denoising network within the Diffusion Model framework enables it to refine the candidate addresses through a process of iterative refinement. 

Considering the locality characteristic of IPv6 address segments and the top-down hierarchical allocation pattern of IPv6, we design a Global-Local Fusion Multi-Head Self-Attention (GLF-MSA) module to capture both the global structure of the IPv6 address space and the local features that are indicative of active addresses, thereby generating high-quality candidate addresses. 6Diffusion can more accurately fit the distribution function of active addresses across the entire IPv6 address space, resulting in the generation of higher-quality candidate addresses.

\textbf{Contributions.} Our contributions can be summarized as follows:
\begin{itemize}
    \item We propose a novel architecture based on diffusion models, termed 6Diffusion, which is capable of fitting the distribution patterns of active addresses, thereby generating more and new active IPv6 addresses.6Diffusion is capable of generating a greater number of new IPv6 address prefixes and is able to produce active IPv6 addresses under these new prefixes.
    \item We propose a Global-Local Fusion Multi-Head Self-Attention (GLF-MSA) module that accommodates the local characteristics of IPv6 address segments and the top-down allocation nature, thereby better learning the underlying features of IPv6.
    \item We have improved the quality of the candidate sets to a higher standard, with experiments demonstrating that 6Diffusion outperforms state-of-the-art target generation algorithms on multiple metrics. For example, the hit rate has been improved to 46.73\%, and the generation rate has been improved to 46. 66\%.
\end{itemize}

\textbf{Roadmap.} Section II summarizes the prior researches related to our work. Section III introduces the overall design of 6Diffusion. Section IV shows the evaluation results. Section V concludes the paper.

\section{RELATED WORK}

Prior work on IPv6 target generation falls into three broad categories: (1) probabilistic statistical information-based algorithms, (2) reinforcement learning-based algorithms, and (3) deep learning-based algorithms.
\subsection{Target Generation Algorithms}
\subsubsection{probabilistic statistical information-based algorithms}

Probabilistic statistical information-based algorithms primarily rely on experience and assumptions; they generate target addresses by analyzing the statistical information of IPv6 addresses. 

For instance, in 2015, Ullrich et al. \cite{ullrich2015reconnaissance} proposed a feature-based recursive address generation method, which utilizes the most frequently occurring addresses in the seed sets to generate new addresses. The method generates Interface Identifiers (IIDs) and then scans these IIDs under a specified prefix.Foremski et al. \cite{foremski2016entropy} introduced Entropy/IP, which conducted entropy analysis on address segments and employed clustering analysis to model different address segments, subsequently using Bayesian networks to generate scanning addresses.Murdock et al. \cite{murdock2017target} proposed the 6Gen, which is based on the Hamming distance of known seed sets to perform agglomerative hierarchical clustering (AHC), generating candidate targets in dense address space regions.In contrast to the AHC division of 6Gen, Liu et al. \cite{liu20196tree} proposed the 6Tree, which uses divisive hierarchical clustering (DHC) to partition address vectors, constructing a space tree and scanning from leaf nodes until the total scanning budget is reached.Song et al. \cite{song2020towards} proposed the DET, which combines the ideas of Entropy/IP and 6Tree, using DHC to construct a space tree; unlike 6Tree, DET uses points with the minimum nybbles (hexadecimal characters) entropy as splitting points and dynamically adjusts the scanning direction of the space tree based on scanning results.The 6Forest \cite{yang20226forest} algorithm further optimizes the algorithm for constructing space trees using DHC, employing a maximum-covering splitting criterion for space division and integrating an enhanced isolation forest algorithm to remove outlier addresses, making the partitioned space tree more rational and effectively reducing the waste of scanning budgets.The HMap6 \cite{hou2023search} algorithm utilizes a bidirectional hierarchical clustering method (BHC), combining the advantages of both agglomerative hierarchical clustering (AHC) and divisive hierarchical clustering (DHC) to generate more promising scanning subspaces. Additionally, the HMap6 algorithm optimizes time complexity.

However, probabilistic statistical information-based algorithms are overly reliant on the quality of the seed set, and most of these algorithms can only generate addresses within the address prefix of the seed set, lacking the ability to generate more extensively.

\subsubsection{reinforcement learning-based algorithms}

Reinforcement learning-based algorithms typically employ the principles of reinforcement learning, generating and probing simultaneously, and dynamically adjusting the scanning direction of the IPv6 address space based on the results of each probe. 

The 6Hit \cite{hou20216hit} algorithm introduces an action reward mechanism during the scanning process, utilizing reinforcement learning algorithm to interact the scanner with the actual network and dynamically adjust the scanning direction based on reward feedback. AddrMiner \cite{song2022addrminer} categorizes IPv6 address regions into three types: areas without seeds, areas with a few seeds, and areas with sufficient seeds. In areas with sufficient seeds, it  learns based on the density characteristics of seeds, corrects density biases caused by seed sampling, and dynamically adjusts the direction of target address generation based on feedback from active addresses. The 6Scan \cite{hou20236scan} algorithm uses the DHC approach, encoding the regional identifiers of target addresses into the probing packets and recording regional activities from asynchronously returned replies, which significantly improves scanning efficiency as the packet sending and receiving processes are asynchronous. It decodes the spatial encoding in the returned packets, calculates the number of active addresses in each region, and dynamically adjusts the search direction.

Although reinforcement learning-based algorithms can perform real-time probing in the real world, they exhibit poor diversity in generated addresses and are prone to becoming trapped in specific regions, particularly in alias regions.

\subsubsection{deep learning-based algorithms}

Deep learning-based algorithms often treat IPv6 addresses as language models, conducting semantic analysis to learn the allocation rules of active IPv6 addresses, thereby generating active IPv6 addresses. 

Cui et al. proposed the 6GCVAE \cite{cui20206gcvae}, which was the first attempt to apply deep learning to the design of IPv6 target generation algorithms. They constructed a Variational Autoencoder (VAE) model to generate active IPv6 addresses. Cui et al. also proposed the 6VecLM \cite{cui20216veclm} algorithm, which constructs the semantic space of IPv6, uses Word2Vec to generate the language vector space of IPv6 addresses, and employs a Transformer network model for training to generate similar IPv6 addresses. Cui et al. proposed 6GAN \cite{cui20216gan}, which attempts to use adversarial neural networks and reinforcement learning to generate addresses. It includes multiple category generators to produce IPv6 addresses and uses an alias discriminator to determine whether an IPv6 address is an alias address.

\section{6DIFFUSION DESIGN}

We first present the system overview of 6Diffusion, As shown in Figure 1, and then describe the details of its key technical components:  1) \textbf{IPv6 address preprocessing}, standardizing IPv6 address formats to facilitate model learning; 2) \textbf{diffusion model training}, utilizing seed sets to learn the characteristics of the active IPv6 address space; 3) \textbf{IPv6 address Generation}, using the trained model to generate candidate set addresses; 4)\textbf{IPv6 address postprocessing}, removing alias addresses present in the candidate set.

\subsection{System Overview}

6Diffusion is a generative algorithm architecture based on diffusion models that can precisely fit the distribution patterns of active addresses through forward process and reverse process. During the forward process, random noise is continuously added to approximate the distribution pattern of the IPv6 seed sets to a normal distribution. Subsequently, a denoising network model is used to progressively denoise the randomly sampled samples, accurately fitting the distribution function of the seed sets IPv6 addresses, ultimately generating potential active IPv6 candidate sets.Additionally, the 6Diffusion algorithm incorporates a Global-Local Fusion Multi-Head Self-Attention (GLF-MSA) module to learn the local characteristics of IPv6 address segments, enabling the generation of new active IPv6 addresses and significantly enhancing the quality of the candidate set addresses, as well as effectively increasing the number of new address prefixes.  

\begin{figure*}[t]
    \centering
    \includegraphics[width=1\textwidth]{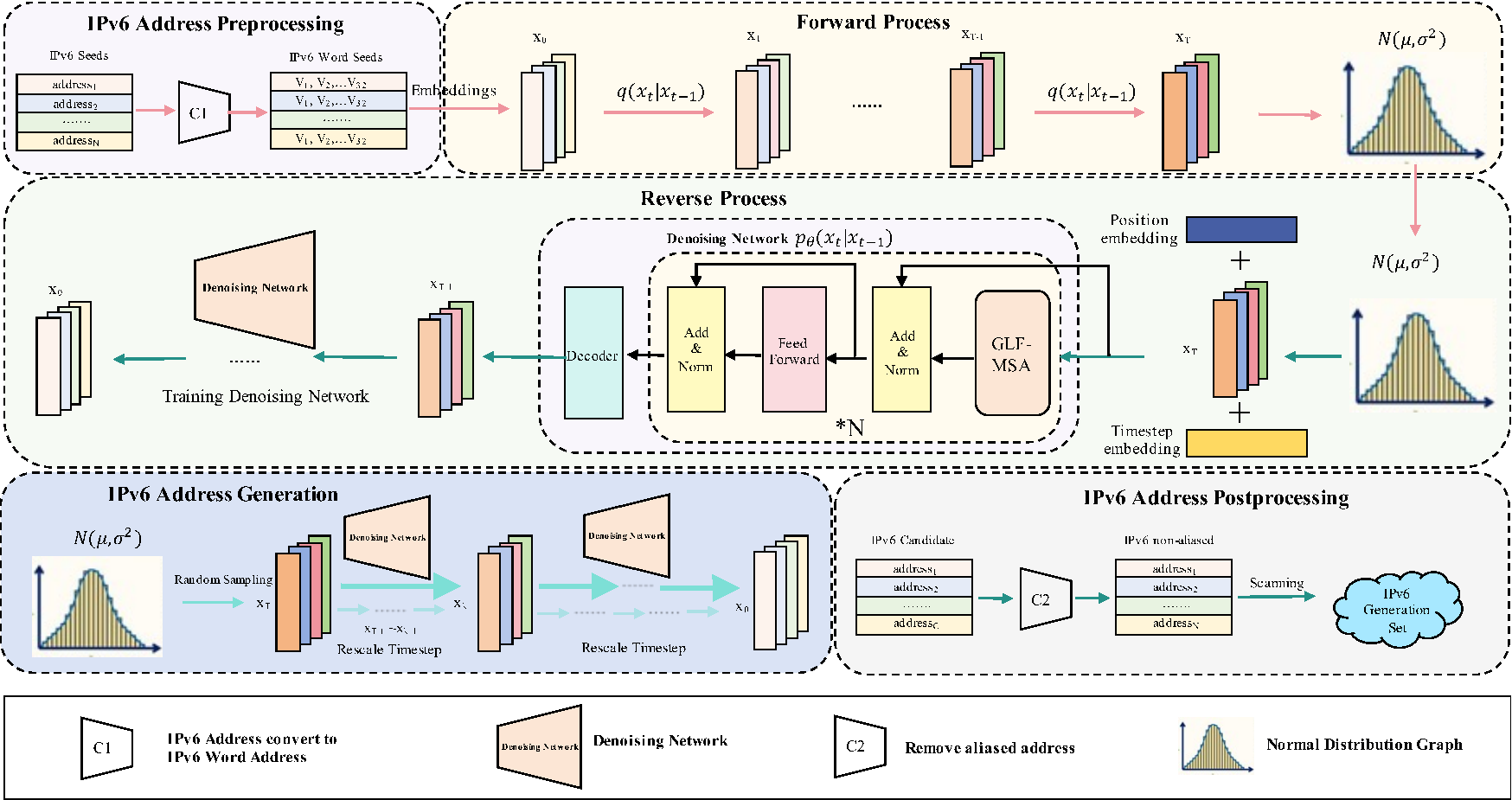}
    \caption{The overall architecture of 6Diffusion, encompassing five key steps: IPv6 Preprocessing, Diffusion Model Training (Forward Process, Reverse Process), Generating IPv6, IPv6 Postprocessing.}
\end{figure*}

\subsection{IPv6 Address Preprocessing}

First, we need to preprocess the IPv6 seed addresses. We consider the obtained active IPv6 seed set as $S=a_1,a_2,\ldots,a_i,\ldots,a_n$, where $a_i, i\in\left[1,\ n\right]$, represents a specific IPv6 address. The IPv6 seed sets S is the input for the entire algorithm, and n denotes the size of the IPv6 seed sets.

To obtain a higher quality seed sets, we need to perform scanning and probing on the input seed sets to identify active addresses within the current seed sets. Then we convert these active addresses into IPv6 word addresses. Since the majority of IPv6 address formats use hexadecimal representation and network administrators are likely accustomed to allocating IPv6 addresses in hexadecimal format, this paper utilizes hexadecimal to represent IPv6 addresses. Given that IPv6 addresses are 128 bits long, they are represented by 32 nybbles in hexadecimal notation. The purpose of IPv6 address preprocessing is to standardize the input format for the convenience of subsequent model training. By conducting prescanning, we can acquire a higher quality set of seed addresses, which allows for the learning of superior characteristics of active addresses.

After representing the active IPv6 addresses in hexadecimal format, we proceed to construct the IPv6 word addresses. We transform the active IPv6 address $a_i$ into an IPv6 word address, where $a_i=\{V_1,V_2,\ldots,V_j,\ldots,V_{32}\}$, $j\in(1,2,\ldots,32)$, Here, $V_j$ denotes the value at the j-th position of the IPv6 address, and $V_j$ can be one of the hexadecimal digits $V_j\ =\{0, 1, 2, 3, 4, 5, 6, 7, 8, 9, a, b, c, d, e, f\}$, The values are separated by spaces.

As shown in Figure 2, the IPv6 address $"2001:0db8:85a3:0000:0000:8a2e:0370:7334"$, after being converted into an IPv6 word address, becomes "2 0 0 1 0 d b 8 8 5 a 3 0 0 0 0 0 0 0 0 8 a 2 e 0 3 7 0 7 3 3 4".

\begin{figure}[H]
    \centering
    \includegraphics[width=0.4\textwidth]{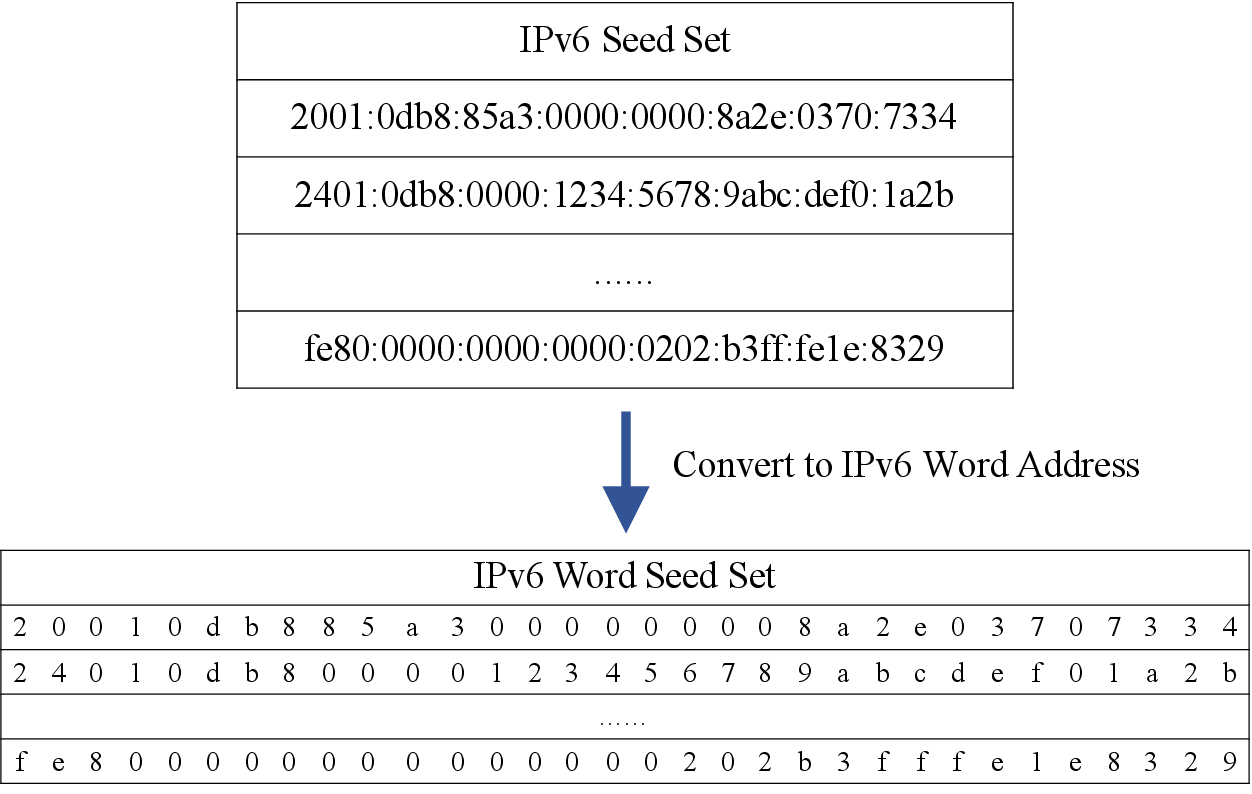}
    \caption{IPv6 Address convert to IPv6 Word Address.}
\end{figure}

\subsection{Diffusion Model Training}

The diffusion model within our architectural framework encompasses two core processes: forward process and reverse process. The 6Diffusion model utilizes the Transformer Network and incorporates a Global-Local Fusion Multi-Head Self-Attention (GLF-MSA) to focus on the local characteristics of IPv6 address segments \cite{hinden2006rfc}, as well as to accommodate the top-down allocation features of IPv6, thereby deeply mining the semantic information of IPv6 addresses and extracting more profound features. The forward process of 6Diffusion involves the gradual introduction of random noise, which is analogous to diffusing the IPv6 seed sets addresses throughout the entire IPv6 address space, aiding the model in more extensively exploring the entire IPv6 address space during the generation phase.

\subsubsection{\textbf{Forward Process}}
The forward process is a process that continuously adds noise to IPv6 address, which can be denoted as $q\left(x_t\middle| x_{t-1}\right)$. For instance, after embedding, we obtain the continuous space IPv6 word vector $x_0$. In the forward process, we introduce Gaussian noise, and over $T$ time steps, we superimpose noise on $x_{t-1}$ at each step according to the following formula:

\begin{equation}
q(x_t \mid x_{t-1}) = \mathcal{N}\left(x_t; \sqrt{1 - \beta_t} x_{t-1}, \beta_t I\right)
\end{equation}

In this context, $\mathcal{N}$ represents Gaussian noise, $I$ denotes the variance of the standard Gaussian noise, $\left\{ \beta_t \in (0, 1) \right\}_{t=1}^T$ indicates the noise schedule that determines the intensity of the noise, and $0 < \beta_1 < \beta_2 < ... <\beta_T$.

After $T$ time steps of diffusion, the continuous space IPv6 word vector $x_0$ will become random noise that follows a Gaussian distribution, as if the active seed IPv6 addresses have been diffused throughout the entire IPv6 address space. Each step of the forward process generates a noisy $x_t$ and the entire forward process can be represented as a Markov chain.

\begin{equation}
q(x_{1:T} \mid x_0) = \prod_{t=1}^{T} q(x_t \mid x_{t-1})
\end{equation}

In the actual forward process, it is possible to directly sample $x_t$ at any step $t$ based on the original data $x_0$, obtaining $x_t\sim q\left(x_t\middle| x_0\right)$. Let $\alpha_t=1-\beta_t$ and $\overline{\alpha_t}=\prod_{i=1}^{t}\alpha_i$. Using the reparameterization trick allows for the completion of the forward process with noise addition in a single computation, without the need for gradual noise addition, and is applicable to any time step $t$.

In this paper, the 6Diffusion model employs linear noise scheduling, with a time step $T = 2000$ steps, and $\beta_1={10}^{-6}, \beta_T=0.01$.

\subsubsection{\textbf{Reverse Process}}
The reverse process can be represented by the conditional probability distribution $p\left(x_{t-1}\middle| x_t\right)$, which is essentially a gradual denoising process and a key step in generating IPv6 addresses. 

The forward process ultimately generates a random noise that follows a Gaussian distribution. The starting point of the reverse process can be a sample from this Gaussian noise, $x_T~\mathcal{N}\left(0,I\right)$. By gradually reducing the noise, we are ultimately able to recover the active IPv6 addresses.

But the conditional probability distribution $p\left(x_{t-1}\middle| x_t\right)$ is not known a priori and must be learned through a denoising network. The denoising network architecture employed in this paper is based on the Transformer, denoted by parameters $\theta$, which learns the conditional probability distribution of $x_{t-1}$ given $x_{t-1}$, i.e., $p_\theta\left(x_{t-1}\middle| x_t\right)$. The noise removed during the reverse process is also Gaussian noise, characterized by its mean and variance. Therefore, $p_\theta\left(x_{t-1}\middle| x_t\right)$ can be derived from the following formula:

\begin{equation}
p_\theta(x_{t-1} \mid x_t) = \mathcal{N}(x_{t-1}; \mu_\theta(x_t, t), \Sigma_\theta(x_t, t))
\end{equation}

The mean $\mu_\theta$ and variance $\Sigma_\theta$ are predicted by the denoising network integrating all time steps.The denoising network is trained to generate an approximation of $x_0$ given $x_t$ and $t$ (timestep). By propagating forward through the denoising network and sampling, and comparing the output vector with the input vector, it is possible to associate the weights of each time step with $x_0$.

When given $x_t$, the primary objective of the predictive network is to estimate the conditional probability distribution $p\left(x_{t-1}\middle| x_t\right)$. With the acquisition of $x_0$, these reverse process probabilities, according to Bayes' theorem, can be expressed in terms of the forward process:

\begin{equation}
p(x_{t-1} \mid x_t, x_0) = \frac{q(x_t \mid x_{t-1}, x_0) q(x_{t-1} \mid x_0)}{q(x_t \mid x_0)}
\end{equation}

Training the generative distribution is equivalent to minimizing the negative log-likelihood of the true samples. This can be indirectly achieved by minimizing the Evidence Lower Bound (ELBO):

\begin{equation}
\begin{aligned}
L_{vb} &= E_{q(x_0)}\left[D_{KL}\left[q(x_T \mid x_0) \mid p(x_T)\right]\right] \\
&\quad + \sum_{t=2}^{T} E_{q(x_t \mid x_0)}\left[D_{KL}\left[p(x_{t-1} \mid x_t, x_0) \mid p_\theta(x_{t-1} \mid x_t)\right]\right] \\
&\quad - E_{q(x_1 \mid x_0)}\left[\log p_\theta(x_1 \mid x_0)\right]
\end{aligned}
\end{equation}

The first term represents the difference between the distribution of the true IPv6 word vector $x_T$ and the standard normal distribution. Since the variance is fixed, this term is a constant and does not need to be optimized during the training process. 

The second term is the KL divergence error of the denoising network, which is mainly used for training the denoising network. The purpose of the second term's loss function is to let the denoising network model learn how to predict noise. By minimizing the difference between the noise introduced during the forward process and the noise predicted by the denoising network model, the model parameters are updated, allowing the learned distribution to be closer to the true distribution. 

The third term is the loss function for mapping back from the continuous space of IPv6 to the discrete space, i.e., the error between $x_0$ and the input of the denoising model, where the mean squared error is used to measure this. Therefore, the loss function for training the 6Diffusion algorithm is the sum of the second and third terms.

\subsubsection{\textbf{Denoising Network}}

To fully capture the semantic features of IPv6 and identify dependencies between different addresses, it is crucial to employ a denoising network model that can capture a variety of dependency relationships. The Unet backbone network, which is based on convolutional networks, can only capture local information through a limited receptive field, limiting its understanding of global context. Consequently, this study selects Transformer as the backbone network to more effectively acquire global information, which aids the model in comprehensively considering the context of the entire sequence when processing sequential data. Given the significant local variations in each address segment of IPv6 and its top-down allocation strategy, we have designed a Global-Local Fusion Multi-Head Self-Attention (GLF-MSA) mechanism to better learn the strategic features of IPv6 allocation.

Initially, we transform the time step t into a format that matches the dimension of the IPv6 address vector, and also encode the positional information as word vectors to be added to it. To explicitly handle specific time steps, we encode the time step as a word vector and incorporate it as well. Ultimately, the model's input is the cumulative result of these three vectors.

\begin{equation}
\text{embedding}_{\text{input}_t} = \text{emb}_{x_t} + \text{emb}_{\text{positon}} + \text{emb}_{\text{timestep}}
\end{equation}

\begin{figure}[H]
    \centering
    \includegraphics[width=0.45\textwidth]{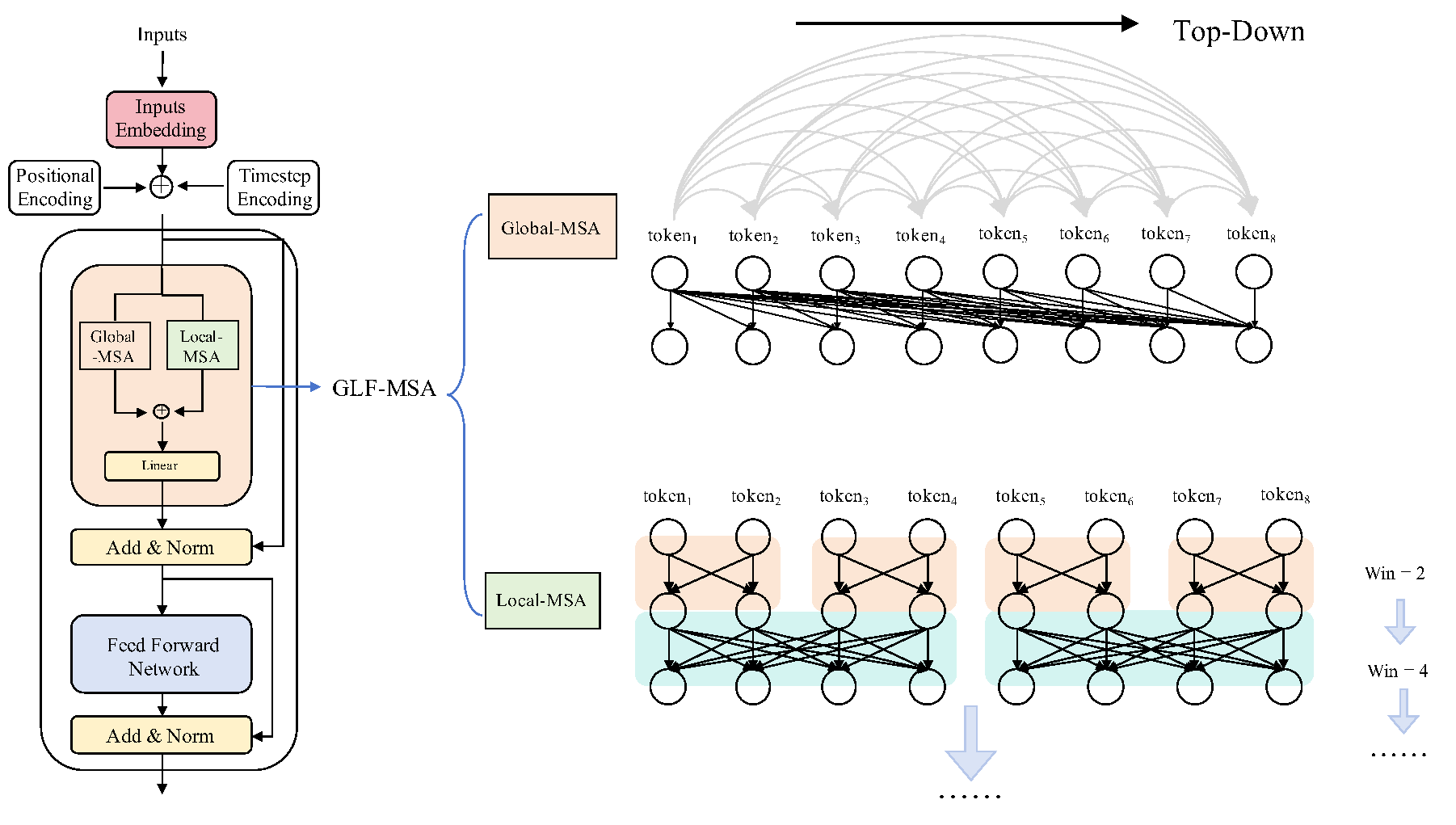}
    \caption{The GLF-MSA module comprises both Global MSA and Local MSA. The Global MSA operates in a top-down manner attention, while the Local MSA is based on multi-scale hierarchical window attention.}
\end{figure}

As shown in Figure 3,the GLF-MSA module comprises both Global MSA and Local MSA. The Global MSA operates in a top-down manner attention, while the Local MSA is based on multi-scale hierarchical window attention. 

Taking into account the top-down allocation approach and the local characteristics within IPv6 address segments of IPv6, we have designed a GLF-MSA module that employs a top-down global multi-head self-attention mechanism and a multi-scale hierarchical window multi-head self-attention mechanism for learning IPv6 addresses. This module consists of two components:Top-down Global Multi-Head Self-Attention and Local Multi-Scale Hierarchical Window Self-Attention.

Top-down Global Multi-Head Self-Attention Mechanism \cite{yang2021causal}: Considering that the IPv6 address allocation pattern is distributed in a top-down manner, the allocation of the current address nybble is not influenced by subsequent address nybble. Therefore, when calculating self-attention, we only take into account the higher-level address information within the current address nybble and exclude the influence of subsequent address nybble. For example, for the third address nybble position, we only compute the attention over the address nybble positions 1-3, and address nybbles at later positions do not affect the global attention at this address nybble position.

Local Multi-Scale Hierarchical Window Self-Attention: Given the various possible local size ranges within IPv6 address segments, we have designed a multi-scale hierarchical window attention mechanism \cite{liu2021swin, guo2020multi}. For example, in the first layer of GLF-MSA, the window size is set to 2, and in the second layer, it is set to 4, and so on, with the window size increasing progressively, akin to a pyramid shape \cite{liu2022pyraformer}, continually expanding the scope of local feature capture. This approach accommodates the local characteristics of IPv6 address segments of different lengths. In this paper, we utilize a 10-layer Transformer Encoder, within which the GLF-MSA's Local-MSA module has varying window sizes for different layers. Specifically, the window size doubles every two layers, starting from 2 in the first two layers and increasing to 32 in the last two layers. These window sizes correspond to the 32 hexadecimal positions of an IPv6 address.

In IPv6 addressing, while local characteristics are important, the allocation of IPv6 addresses is also influenced by global factors. The global attention mechanism can capture long-range dependencies in sequential data, which is crucial for understanding the global structure and overall patterns of IPv6 address allocation. Therefore, we integrate the global attention module with the local attention module, the GLF-MSA, enabling the model to learn more features of IPv6 and generate higher-quality candidate sets.

\subsection{IPv6 Address Generation}

After the model training is completed, we utilize the trained 6Diffusion model to generate IPv6 addresses, allowing for customization of the number of addresses generated based on requirements. To generate $M$ candidate addresses, we randomly sample $M$ samples from a normal distribution and then perform multiple denoising processes through the denoising network, ultimately obtaining several generated IPv6 addresses that form the IPv6 candidate sets. Thanks to the trained diffusion model's ability to effectively learn the semantic information of active IPv6 addresses, the model's generation process can accurately identify potential active IPv6 addresses $x_0$ even after random sampling of $x_T$. Therefore, the model proposed in this study is a streamlined denoising network based on diffusion models, capable of achieving end-to-end IPv6 address generation without the need for a complex multi-stage processing workflow.

To accelerate the generation of IPv6 addresses, we employ the DDIM \cite{song2020denoising} sampling acceleration method, which utilizes a skip-step approach to expedite sampling and enhance the speed of IPv6 address generation. Assuming we have data $x_t$ at step $t$, along with the predicted noise $\epsilon_\theta\left(x_t,t\right)$, the forward process can be approximated using the formula as follows:

\begin{equation}
    x_t = \sqrt{\alpha_t} x_0 + \sqrt{1 - \alpha_t} \epsilon
\end{equation}

The reverse process involves recovering from $x_t$ to $x_{t-1}$. Initially, we estimate $x_0$ to obtain:

\begin{equation}
    \widehat{x_0} = \frac{x_t - \sqrt{1 - \alpha_t} \epsilon_\theta(x_t, t)}{\sqrt{\alpha_t}}
\end{equation}

Subsequently, we utilize the estimated estimate $\widehat{x_0}$ to calculate $x_{t-1}$:

\begin{equation}
    x_{t-1}=\sqrt{\alpha_{t-1}}\widehat{x_0}+\sqrt{{1-\alpha}_{t-1}}\epsilon_\theta\left(x_t,t\right)
\end{equation}

From Equations 1 and 12, we can derive that:

\begin{equation}
    x_{t-1}=\sqrt{\alpha_{t-1}}\frac{x_t-\sqrt{1-\alpha_t}\epsilon_\theta\left(x_t,t\right)}{\sqrt{\alpha_t}}+\sqrt{{1-\alpha}_{t-1}}\epsilon_\theta\left(x_t,t\right)
\end{equation}

Through a series of derivations and utilizing the DDIM (Denoising Diffusion Implicit Model) sampling acceleration method, we have achieved a deterministic reverse process from noise data to original data, avoiding the randomness of sampling at each step. This significantly enhances the efficiency of IPv6 address generation while retaining the high quality of the generated data.

\begin{algorithm}
\caption{Sampling Process to Generate IPv6}
\begin{algorithmic}[1]
\State \textbf{input:} Random Sampling $x_T \sim \mathcal{N}(0, I)$
\State \textbf{Initialize Timesteps:} Let $t_K = T$ and $t_1 = 0$
\State \textbf{Rescale Timesteps:} Define the sequence of timesteps \newline
$t = t_K, t_{K-1}, \ldots, t_2, t_1$
\For{\text{Timesteps: }$t_k = K, \ldots, 1$}
    \State \textbf{Predict noise:} $\epsilon_{\theta}(x_{t_k}, t_k)$
    \State \textbf{Estimate } $\hat{x}_0$:
    \begin{equation*}
    \hat{x}_0 = \frac{x_{t_k} - \sqrt{1 - \alpha_{t_k}} \epsilon_{\theta}(x_{t_k}, t_k)}{\sqrt{\alpha_{t_k}}}
    \end{equation*}
    \State \textbf{Update data:} $x_{t_{k-1}}$
    \begin{equation*}
    x_{t_{k-1}} = \sqrt{\alpha_{t_{k-1}}} \hat{x}_0 + \sqrt{1 - \alpha_{t_{k-1}}} \epsilon_{\theta}(x_{t_k}, t_k)
    \end{equation*}
\EndFor
\State \textbf{return} $\text{IPv6 address }x_0$
\end{algorithmic}
\end{algorithm}

\subsection{IPv6 Address Postprocessing}

\begin{figure}[H]
    \centering
    \includegraphics[width=0.45\textwidth]{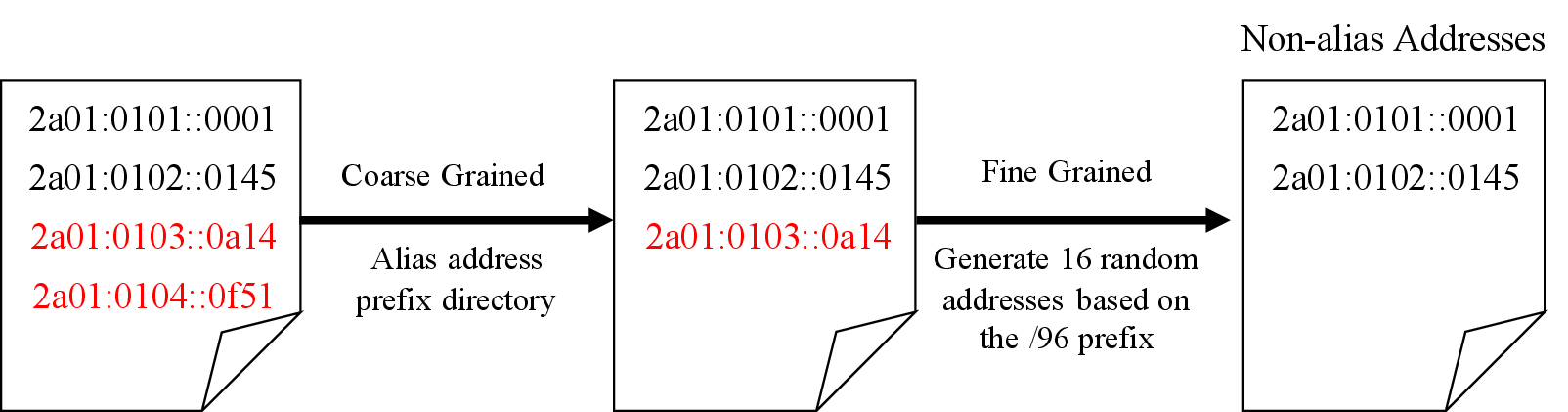}
    \caption{Remove alias address in two steps: coarse- grained and fine-grained.}
\end{figure}

After obtaining the candidate address set C, we need to further perform deduplication and alias address cleanup operations on the C. The alias removal method in this study is divided into two stages: First, we use known alias address prefixes to perform a preliminary coarse-grained alias removal on the candidate sets, thereby filtering out a set of non-alias addresses. This step can quickly and efficiently identify alias addresses. However, this method largely depends on the collected alias address prefixes, as shown in Figure 4, so we conduct a more refined alias removal operation on the address set that has undergone coarse-grained processing.

For each IPv6 address that has undergone coarse-grained processing, we perform the following operations: fix the first 96 bits as the prefix and randomly generate 16 distinct addresses for the next 32 bits (i.e., bits 97-128), ensuring that the first 96 bits of these 16 addresses remain the same while the remaining 32 bits are randomly generated. Subsequently, we conduct probing scans on these 16 addresses. If all 16 addresses are found to be active, the address is determined to be an alias and is excluded; if not all are active, it is considered a non-alias address, preserved for further analysis.

\section{PERFORMANCE EVALUATION}
In this section, we present our experimental setup and all experimental results to indicate 6Diffusion’s performance.

\subsection{Experimental Setup}
\subsubsection{Dataset}
Gasser et al. \cite{gasser2018clusters} have provided a public dataset known as the IPv6 Hitlist, which aims to facilitate the development of future IPv6 research. This dataset compiles daily scanning results from various public address sets and employs scanning strategies to achieve daily updates of alias prefix data. In this study, we utilized the dataset from May 10, 2024, which contained over twenty million active addresses.

Additionally, our laboratory has conducted active scanning and probing tasks on the real-world IPv6 address space, and the collected active IPv6 seed set has reached over five million. Therefore, we randomly selected 100k of these addresses as the experimental seed set.
\subsubsection{Environment}
The development platform for this algorithm is Python 3.8.10 and PyTorch 1.10.0, with the software environment being Ubuntu 20.04.3. The hardware environment consists of a CPU: Intel(R) Xeon(R) Platinum 8352V CPU @ 2.10GHz, 32G of memory, and an RTX 4090 GPU with 24G of video memory.

To ensure the accuracy of the results, we designed a series of scanning operations to verify the active addresses generated by the target generation algorithms. We employed the Zmapv6 tool \cite{durumeric2013zmap}, utilizing a variety of protocols including ICMPv6, TCP/80, TCP/443, UDP/53, and UDP/443 for comprehensive scanning. If any of the scanning protocols detect a response, the address is marked as an active target.

\subsubsection{Model Parameter}
During the training process of the diffusion model, we first convert each word in the IPv6 word address into a 64-dimensional vector. We utilize a Transformer architecture with 10 Encoder layers, each with 512 hidden units. The Global-MSA module in the GLF-MSA employs 2 attention heads, while the Local-MSA module in the GLF-MSA uses different window sizes for different layers. Specifically, the window size doubles every two layers, starting from 2 in the first two layers and increasing to 32 in the last two layers, and every Local-MSA layer employs 2 attention heads each. After concatenation, a linear layer is used to transform the vectors back to their original dimensionality. The diffusion step is set to 2000, with the noise schedule employing a linear schedule, $\beta_1={10}^{-6} and \beta_{2000}=0.01$. The batch size is 512, the dropout rate is 0.1, and the learning rate is 0.001.

During the generation of IPv6 addresses, we rescale the diffusion steps, sampling every 5 steps, to accelerate the speed of IPv6 address generation.

\subsubsection{Evaluation Metrics}
In terms of evaluation metrics for this study, we have adopted standard metrics used in IPv6 target generation algorithms: hit rate, generation rate, and non-alias rate. Additionally, this study introduces two innovative evaluation metrics: candidate new prefix rate and generation new prefix rate. These new metrics aim to measure the diversity of addresses in the candidate sets and the generated sets, thereby assessing the distribution of the generated addresses.

\begin{itemize}
    \item \textbf{Hit rate.}The hit rate refers to the proportion of active addresses in the IPv6 candidate set, which is the ratio of the number of active addresses to the total number of addresses in the candidate set. This metric measures the model's ability to learn from seed addresses.
    \begin{equation}
        r_{hit}=\frac{N_{hit}}{N_{candidate}}
    \end{equation}
    
    \item \textbf{Generation rate.}The generation rate refers to the proportion of newly discovered active IPv6 addresses in the candidate set to the total number of addresses in the candidate set. This metric validates the model's capability to generate new active addresses within the IPv6 address space. $N_{repeat}$ represents the number of addresses in the candidate set that are identical to the addresses in the seed set.
    \begin{equation}
        r_{gen}=\frac{N_{hit}-N_{repeat}}{N_{candidate}}
    \end{equation}
    
    \item \textbf{Non-Alias rate.}The non-alias rate refers to the proportion of non-alias addresses in the candidate set, which is the ratio of the number of non-alias addresses to the total number of addresses in the candidate set. This metric measures the model's ability to generate non-alias addresses within the IPv6 address space.
    \begin{equation}
        r_{non-aliased}=\frac{N_{candidate}-N_{aliased}}{N_{candidate}}
    \end{equation}

    \item \textbf{Candidate new prefix rate.}The candidate new prefix rate refers to the proportion of newly generated address prefixes in the candidate set to the total number of addresses in the candidate set. This metric verifies the model's ability to discover new prefixes within the IPv6 address space, rather than being limited to the prefixes of seed addresses. The significance of this metric is the number of new address prefixes discovered per candidate address, which can validate the diversity of addresses generated by the algorithm. $Set\left(C\right)_{pre}$ represents the number of candidate address prefix sets. $Set\left(S\right)_{pre}$ represents the number of seed prefix address sets.
    \begin{equation}
        r_{cn-pre}=\frac{Num\left(Set\left(C\right)_{pre}-Set\left(S\right)_{pre}\right)}{N_{candidate}}
    \end{equation}

    \item \textbf{Generation new prefix rate.}The generation set new prefix rate refers to the proportion of newly active address prefixes in the generation set to the total number of addresses in the candidate set. This metric validates the model's ability to identify new active prefixes within the IPv6 address space. The numerical significance of this metric is the number of new active address prefixes discovered per candidate address, which can verify the diversity of newly generated active addresses by the algorithm. $Set\left(G\right)_{pre}$ represents the number of generation prefix address sets.
    \begin{equation}
        r_{gn-pre}=\frac{Num\left(Set\left(G\right)_{pre}-Set\left(S\right)_{pre}\right)}{N_{candidate}}
    \end{equation}
    
\end{itemize}

\subsection{Target Generation Algorithms Performance}
We compared the performance of 6Diffusion with other state-of-the-art target generation algorithms, including Entropy/IP, 6Tree, DET, 6Hit, 6GCVAE, 6VecLM, 6Forest, 6GAN, AddrMiner, HMap6, and 6Scan as baselines. In the above dataset, we not only compared the common performance metrics of TGAs (hit rate and generation rate) but also considered the quality of the candidate set from various angles and proposed multiple metrics to assess the quality of the generated addresses. This demonstrates that 6Diffusion outperforms previous studies in terms of performance excellence.

\begin{table}[H]
    \centering
\caption{Comparison of Different Target Algorithm Performances with 100k Seeds}
     \scalebox{0.7}{
    \begin{tabular}{cccccccc} 
        \toprule
        \textbf{Method} & \textbf{$N_{\text{candidate}}$} & \textbf{$N_{\text{nonaliased}}$} & \textbf{$r_{\text{nonaliased}}$} & \textbf{$N_{\text{hit}}$} & \textbf{$N_{\text{gen}}$} & \textbf{$r_{\text{hit}}$} & \textbf{$r_{\text{gen}}$} \\
        \midrule
        Entropy/IP\cite{foremski2016entropy} & 100000 & 91848 & 91.85\% & 5130 & 5130 & 5.13\% & 5.13\% \\
        \addlinespace
        6Tree\cite{liu20196tree} & 100000 & 50507 & 50.51\% & 25199 & 22403 & 25.20\% & 22.40\% \\
        \addlinespace
        DET\cite{song2020towards} & 100000 & 52921 & 52.92\% & 27744 & 24600 & 27.74\% & 24.60\% \\
        \addlinespace
        6Hit\cite{hou20216hit} & 136874 & 12707 & 9.28\% & 8312 & 8095 & 6.07\% & 5.91\% \\
        \addlinespace
        6GCVAE\cite{cui20206gcvae} & 100000 & 17874 & 17.87\% & 1848 & 1845 & 1.85\% & 1.85\% \\
        \addlinespace
        6VecLM\cite{cui20216veclm} & 173555 & 102177 & 58.87\% & 37838 & 11223 & 21.80\% & 6.47\% \\
        \addlinespace
        6Forest\cite{yang20226forest} & 106904 & 67334 & 62.99\% & 25043 & 24683 & 23.43\% & 23.09\% \\
        \addlinespace
        6GAN\cite{cui20216gan} & 208192 & 151857 & 72.94\% & 37176 & 37083 & 17.86\% & 17.81\% \\
        \addlinespace
        AddrMiner\cite{song2022addrminer} & 100000 & 70933 & 70.93\% & 40679 & 36682 & 40.68\% & 36.68\% \\
        \addlinespace
        6HMap\cite{hou2023search} & 100000 & 75761 & 75.76\% & 41446 & 39060 & 41.45\% & 39.06\% \\
        \addlinespace
        6Scan\cite{hou20236scan} & 100080 & 25535 & 25.51\% & 16956 & 16825 & 16.94\% & 16.81\% \\
        \bottomrule
        \addlinespace
        6Diffusion & 95092 & 88528 & \textbf{93.10\%} & \textbf{44435} & \textbf{44372} & \textbf{46.73\%} & \textbf{46.66\%} \\
        \bottomrule
    \end{tabular}
    }
    
\end{table}

\begin{figure}[H]
    \centering
    \includegraphics[width=0.49\textwidth]{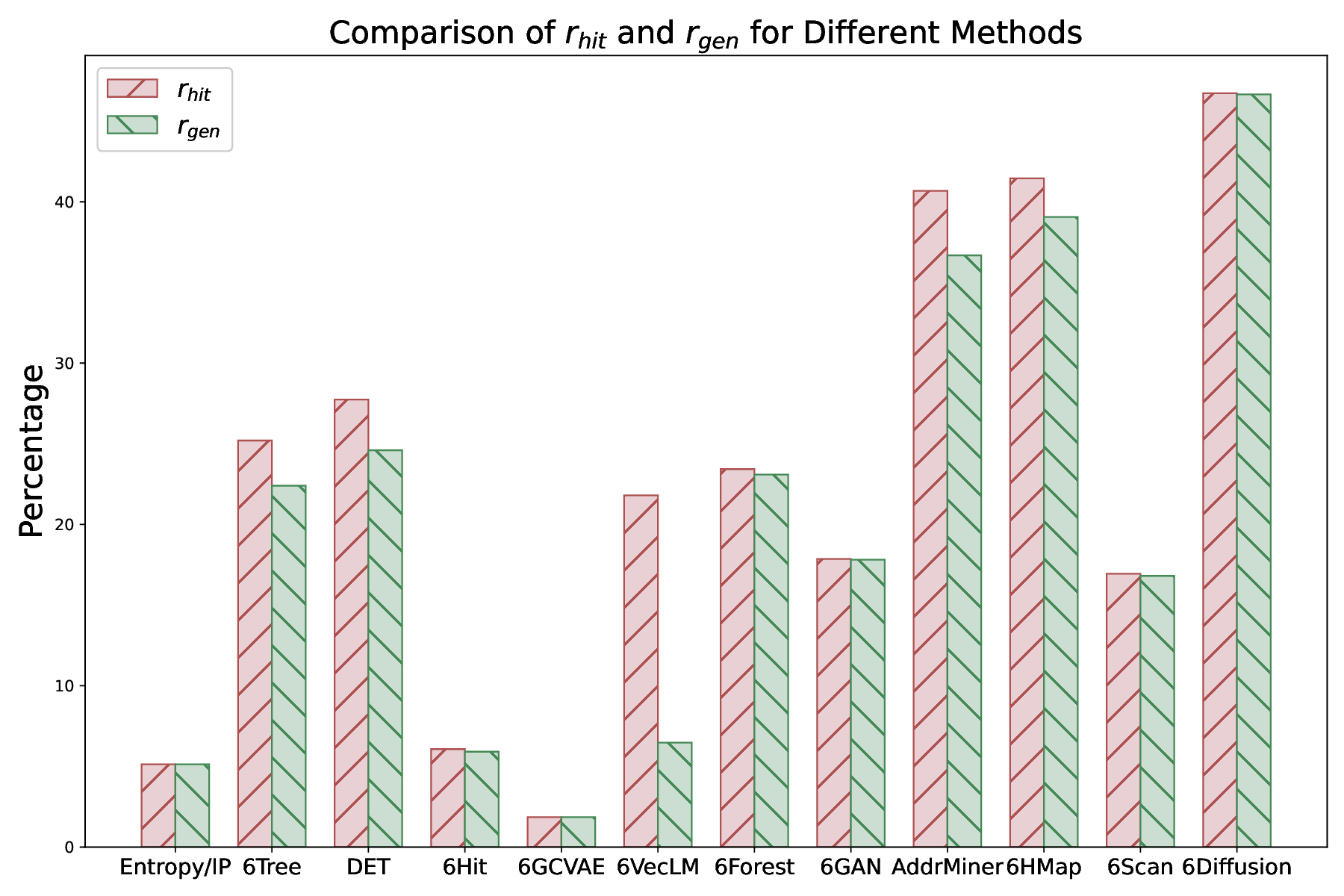}
    \caption{Hit rate and Generation rate of target generation algorithms.}
\end{figure}

\subsubsection{\textbf{Hit rate and Generation rate Performance}}
As shown in Table I and Figure 5, regarding the hit rate metric, methods such as 6Tree, DET, 6VecLM, 6Forest, 6GAN, and 6Scan exhibit average performance, with hit rates ranging from 18.07\% to 28.94\%. In contrast, AddrMiner and HMap6 demonstrate better performance, with hit rates between 33.28\% and 38.70\%.

For the generation rate metric, Entropy/IP and 6Hit perform poorly. However, the 6GCVAE algorithm, which has a good hit rate, performs poorly in terms of generation rate, indicating that most of the candidate IPv6 addresses are already present in the seed set. The performance of 6Tree, DET, 6VecLM, 6Forest, 6GAN, AddrMiner, HMap6, and 6Scan ranges from 17.97\% to 36.31\%.

The 6Diffusion algorithm shows significant improvements in both hit rate and generation rate, with a hit rate of 46.73\% and a generation rate of 46.66\%. Compared to other algorithms, the hit rate improvement ranges from 5.28\% to 44.8\%, and the generation rate improvement ranges from 7.6\% to 44.8\%. This indicates that the 6Diffusion algorithm not only effectively fits the distribution pattern of active addresses in the seed set but also generates a greater number of new active addresses within the IPv6 address space.

\subsubsection{\textbf{Non-Aliased rate Performance}}
As shown in Table 1 and Figure 6, regarding the non-alias rate metric, the Entropy/IP algorithm performs well in terms of non-alias rates, primarily because it generates a majority of non-active addresses, which leads to unsatisfactory performance in hit rate and generation rate. The non-alias rate performance of the 6Hit, 6GCVAE, and 6Scan methods is underwhelming. Particularly, 6Hit and 6Scan, as reinforcement learning-based algorithms, dynamically adjust during scanning and generation, but once they become trapped in alias address regions, they produce a large number of alias addresses, thereby consuming valuable scanning resources. Experimental results confirm this, as 6Hit, 6GCVAE, and 6Scan indeed became trapped in alias regions, generating a significant number of alias addresses.

In contrast, the 6Diffusion algorithm demonstrates superior performance in fitting the distribution pattern of non-alias addresses, resulting in the majority of generated addresses being non-alias, with a non-alias rate reaching 93.10\%. Compared to other methods (excluding Entropy/IP), the improvement of 6Diffusion is significant.

\begin{figure}[H]
    \centering
    \includegraphics[width=0.49\textwidth]{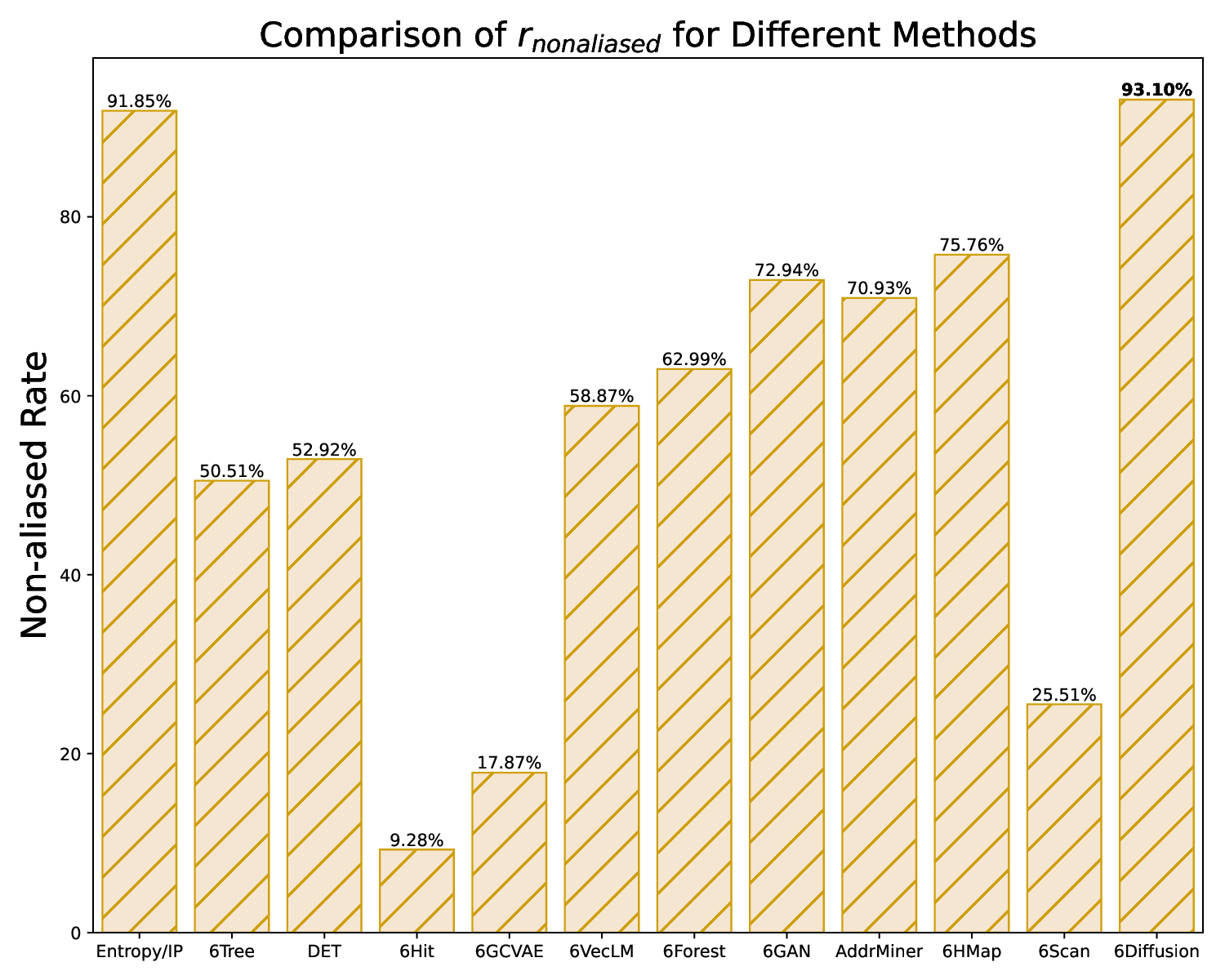}
    \caption{Non-alias rate of target generation algorithms.}
\end{figure}

\subsubsection{\textbf{Candidate new prefix rate and Generation new prefix rate Performance}}

The candidate new prefix rate and generation new prefix rate metrics reflect the algorithm's ability to generate new address prefixes. A higher metric value indicates that the algorithm can generate addresses that are more dispersed, rather than being concentrated under the seed address prefixes, which reflects the diversity of addresses generated by the algorithm.

To better compare the performance of algorithms across different lengths of address prefixes, we compared four different lengths of address prefixes (/32, /48, /64, /80), with the results depicted in Table II-V. The Entropy/IP algorithm generated a large number of new prefixes in the candidate set; however, due to the high randomness of its generated addresses and the resulting low hit rate and generation rate, we decided to exclude it from subsequent comparative analysis. Due to the small values of the candidate new prefix rate and the generation new prefix rate, to facilitate a more intuitive comparison, we multiply both rates by 10,000, representing the number of new prefixes generated per 10,000 candidate addresses. This allows for a more direct perception of the performance differences between various algorithms.

\begin{table}
    \centering
\caption{COMPARISION OF CANDIDATE NEW PREFIX RATE AND GENERATION NEW PREFIX RATE UNDER THE /32 PREFIX}
         \scalebox{0.75}{
    \begin{tabular}{c>{\centering\arraybackslash}p{0.07\linewidth}>{\centering\arraybackslash}p{0.07\linewidth}>{\centering\arraybackslash}p{0.07\linewidth}>{\centering\arraybackslash}p{0.12\linewidth}>{\centering\arraybackslash}p{0.07\linewidth}>{\centering\arraybackslash}p{0.07\linewidth}>{\centering\arraybackslash}p{0.07\linewidth}>{\centering\arraybackslash}p{0.12\linewidth}}
         \multicolumn{9}{c}{/32 address prefix}\\
         \toprule
\textbf{Method}& \textbf{$N_{\text{c-pre}}$}& \textbf{$N_{\text{cn-pre}}$}& \textbf{$  r_{cn-pre}$}& \textbf{$N_{\text{cn-pre}}$ $per_{10k}$}& \textbf{$N_{\text{g-pre}}$}& \textbf{$N_{\text{gn-pre}}$}& \textbf{$  r_{gn-pre}$}& \textbf{$N_{\text{gn-pre}}$ $per_{10k}$}\\
         \midrule
        Entropy/IP\cite{foremski2016entropy} & 42172 & 41019 & 41.02\% & 4101.90 & 17 & 0 & 0.00\% & 0.00 \\
            \addlinespace
        6Tree\cite{liu20196tree} & 3821 & 0 & 0.00\% & 0.00 & 1174 & 0 & 0.00\% & 0.00 \\
        \addlinespace
        DET\cite{song2020towards} & 3837 & 0 & 0.00\% & 0.00 & 1248 & 0 & 0.00\% & 0.00 \\
        \addlinespace
        6Hit\cite{hou20216hit} & 108 & 0 & 0.00\% & 0.00 & 85 & 0 & 0.00\% & 0.00 \\
        \addlinespace
        6GCVAE\cite{cui20206gcvae} & 1084 & 1002 & 1.00\% & 100.20 & 30 & 15 & 0.02\% & 1.50 \\
        \addlinespace
        6VecLM\cite{cui20216veclm} & 4558 & 0 & 0.00\% & 0.00 & 1316 & 0 & 0.00\% & 0.00 \\
        \addlinespace
        6Forest\cite{yang20226forest} & 3993 & 2081 & 1.95\% & 194.66 & 644 & 80 & 0.07\% & 7.48 \\
        \addlinespace
        6GAN\cite{cui20216gan} & 16491 & 14582 & \textbf{7.00\%} & \textbf{700.41} & 80 & 7 & 0.00\% & 0.34 \\
        \addlinespace
        AddrMiner\cite{song2022addrminer} & 421 & 129 & 0.13\% & 12.9 & 180 & 1 & 0.00\% & 0.10 \\
        \addlinespace
        6HMap\cite{hou2023search} & 318 & 16 & 0.02\% & 1.74 & 186 & 2 & 0.00\% & 0.22 \\
        \addlinespace
        6Scan\cite{hou20236scan} & 179 & 0 & 0.00\% & 0.00 & 123 & 0 & 0.00\% & 0.00 \\
        \bottomrule
        \addlinespace
        6Diffusion & 4104 & 3266 & 3.43\% & 343.45 & 310 & 131 & \textbf{0.14\%} & \textbf{13.78} \\
        \bottomrule
    \end{tabular}
    
    }
\end{table}

\textbf{/32 address prefix.} As shown in Table II, under the /32 prefix, the 6GAN algorithm performs the best, generating more new address prefixes, while other algorithms exhibit relatively poor performance, generating approximately 700 new prefixes for every 10k candidate addresses. The 6Diffusion algorithm also demonstrates good performance under the /32 prefix, generating about 343 new prefixes for every 10k candidate addresses.

As illustrated in Table II, regarding the generation of new active address prefixes under the /32 prefix, the 6Diffusion model is capable of generating the highest number of /32 active address prefixes, with 13 active prefixes generated for every 10k candidate addresses, outperforming other algorithms. This indicates that the 6Diffusion algorithm not only generates a more diverse set of address prefixes but also uncovers a greater number of new active prefix addresses.

\begin{table}[H]
    \centering
\caption{COMPARISION OF CANDIDATE NEW PREFIX RATE AND GENERATION NEW PREFIX RATE UNDER THE /48 PREFIX}
         \scalebox{0.75}{
    \begin{tabular}{c>{\centering\arraybackslash}p{0.07\linewidth}>{\centering\arraybackslash}p{0.07\linewidth}>{\centering\arraybackslash}p{0.07\linewidth}>{\centering\arraybackslash}p{0.12\linewidth}>{\centering\arraybackslash}p{0.07\linewidth}>{\centering\arraybackslash}p{0.07\linewidth}>{\centering\arraybackslash}p{0.07\linewidth}>{\centering\arraybackslash}p{0.12\linewidth}}
         \multicolumn{9}{c}{/48 address prefix}\\
         \toprule
\textbf{Method}& \textbf{$N_{\text{c-pre}}$}& \textbf{$N_{\text{cn-pre}}$}& \textbf{$  r_{cn-pre}$}& \textbf{$N_{\text{cn-pre}}$ $per_{10k}$}& \textbf{$N_{\text{g-pre}}$}& \textbf{$N_{\text{gn-pre}}$}& \textbf{$  r_{gn-pre}$}& \textbf{$N_{\text{gn-pre}}$ $per_{10k}$}\\
         \midrule
       Entropy/IP\cite{foremski2016entropy} & 91255 & 89771 & 89.77\% & 8977.10 & 5078 & 4825 & 4.83\% & 482.50 \\
            \addlinespace
         6Tree\cite{liu20196tree} & 32360 & 103 & 0.10\% & 10.30 & 12289 & 47 & 0.05\% & 4.70 \\
         \addlinespace
        DET\cite{song2020towards} & 32549 & 570 & 0.57\% & 57.00 & 13357 & 275 & 0.28\% & 27.50 \\
        \addlinespace
        6Hit\cite{hou20216hit} & 302 & 78 & 0.06\% & 5.70 & 203 & 36 & 0.03\% & 2.63 \\
        \addlinespace
        6GCVAE\cite{cui20206gcvae} & 7616 & 6735 & 6.74\% & 673.50 & 904 & 592 & 0.59\% & 59.20 \\
        \addlinespace
        6VecLM\cite{cui20216veclm} & 42595 & 0 & 0.00\% & 0.00 & 7105 & 0 & 0.00\% & 0.00 \\
        \addlinespace
        6Forest\cite{yang20226forest} & 17554 & 12249 & 11.46\% & 1145.78 & 3894 & 874 & 0.82\% & 81.75 \\
        \addlinespace
        6GAN\cite{cui20216gan} & 109299 & 97779 & \textbf{46.97\%} & \textbf{4696.58} & 24851 & 19320 & \textbf{9.28\%} & \textbf{927.99} \\
        \addlinespace 
        AddrMiner\cite{song2022addrminer} & 15751 & 8407 & 8.44\% & 843.57 & 9783 & 4761 & 4.78\% & 477.72 \\
        \addlinespace
        HMap6\cite{hou2023search}& 4583 & 2821 & 3.07\% & 306.66 & 2704 & 1387 & 1.51\% & 150.78 \\
        \addlinespace
        6Scan\cite{hou20236scan} & 1013 & 470 & 0.52\% & 52.22 & 731 & 341 & 0.38\% & 37.89 \\
        \bottomrule
        \addlinespace
        6Diffusion & 26081 & 21605 & 22.72\% & 2271.99 & 11286 & 7988 & 8.40\% & 840.02 \\
        \bottomrule
    \end{tabular}
    
    }
\end{table}

\textbf{/48 address prefix.} As shown in Table III, under the /48 prefix, the 6GAN algorithm generated the most address prefixes in the candidate set, producing 4696 prefixes for every 10k candidate addresses. The 6Diffusion algorithm from this study also performed well, generating 2271 prefixes for every 10k candidate addresses.

As illustrated in Table III, in terms of generating new active address prefixes under the /48 prefix, the 6GAN algorithm again demonstrated excellent performance, generating 927 active prefixes for every 10k candidate addresses. The 6Diffusion algorithm also showed good results in generating new active address prefixes, producing 840 active prefixes for every 10k candidate addresses. Compared to other algorithms, these two algorithms stand out in their effectiveness.

\begin{table}[H]
    \centering
\caption{COMPARISION OF CANDIDATE NEW PREFIX RATE AND GENERATION NEW PREFIX RATE UNDER THE /64 PREFIX}
         \scalebox{0.75}{
    \begin{tabular}{c>{\centering\arraybackslash}p{0.07\linewidth}>{\centering\arraybackslash}p{0.07\linewidth}>{\centering\arraybackslash}p{0.07\linewidth}>{\centering\arraybackslash}p{0.12\linewidth}>{\centering\arraybackslash}p{0.07\linewidth}>{\centering\arraybackslash}p{0.07\linewidth}>{\centering\arraybackslash}p{0.07\linewidth}>{\centering\arraybackslash}p{0.12\linewidth}}
         \multicolumn{9}{c}{/64 address prefix}\\
         \toprule
\textbf{Method}& \textbf{$N_{\text{c-pre}}$}& \textbf{$N_{\text{cn-pre}}$}& \textbf{$  r_{cn-pre}$}& \textbf{$N_{\text{cn-pre}}$ $per_{10k}$}& \textbf{$N_{\text{g-pre}}$}& \textbf{$N_{\text{gn-pre}}$}& \textbf{$  r_{gn-pre}$}& \textbf{$N_{\text{gn-pre}}$ $per_{10k}$}\\
         \midrule
            Entropy/IP\cite{foremski2016entropy}& 98615 & 98495 & 98.50\% & 9849.50 & 5130 & 5129 & 5.13\% & 512.90 \\
            \addlinespace
            6Tree\cite{liu20196tree} & 60229 & 19348 & 19.35\% & 1934.80 & 22403 & 12362 & 12.36\% & 1236.20 \\
            \addlinespace
            DET\cite{song2020towards} & 62618 & 24878 & 24.88\% & 2487.80 & 24600 & 14860 & 14.86\% & 1486.00 \\
            \addlinespace
            6Hit\cite{hou20216hit} & 12710 & 12355 & 9.03\% & 902.65 & 8095 & 8020 & 5.86\% & 585.94 \\
            \addlinespace
            6GCVAE\cite{cui20206gcvae} & 19181 & 19135 & 19.14\% & 1913.50 & 1845 & 1842 & 1.84\% & 184.20 \\
            \addlinespace
            6VecLM\cite{cui20216veclm} & 73616 & 0 & 0.00\% & 0.00 & 8934 & 0 & 0.00\% & 0.00 \\
            \addlinespace
            6Forest\cite{yang20226forest} & 67425 & 65491 & 61.26\% & 6126.09 & 24683 & 24266 & 22.70\% & 2269.87 \\
            \addlinespace
            6GAN\cite{cui20216gan} & 182824 & 179414 & 86.18\% & 8617.72 & 37083 & 36556 & 17.56\% & 1755.88 \\
            \addlinespace
            AddrMiner\cite{song2022addrminer} & 70946 & 63937 & 64.16\% & 6415.51 & 37449 & 35537 & 35.66\% & 3565.82 \\
            \addlinespace
            HMap6\cite{hou2023search} & 75766 & 74240 & 80.70\% & 8070.44 & 39060 & 38947 & 42.34\% & 4233.83 \\
            \addlinespace
            6Scan\cite{hou20236scan} & 25541 & 25060 & 27.84\% & 2784.44 & 16825 & 16618 & 18.46\% & 1846.44 \\
            \bottomrule
            \addlinespace
            6Diffusion & 90194 & 89538 & \textbf{94.16\%} & \textbf{9415.84} & 43807 & 43557 & \textbf{45.80\%} & \textbf{4580.46} \\
        \bottomrule
    \end{tabular}
    
    }
\end{table}

\textbf{/64 address prefix.} As depicted in Table IV, under the /64 prefix, the 6Diffusion algorithm from this study generated the highest number of address prefixes in the candidate set, specifically producing 9415 new prefixes for every 10k candidate addresses. Although most other algorithms also demonstrated decent performance under the /64 prefix, the 6Diffusion algorithm particularly stood out, generating the most address prefixes.

As shown in Table IV, in terms of generating new active address prefixes under the /64 prefix, the 6Diffusion algorithm from this study again performed the best, generating 4580 active address prefixes for every 10k candidate addresses.

These results indicate that 6Diffusion algorithm performs exceptionally well under the /64 prefix, aligning with the structural characteristics of IPv6 addresses. The first 48 bits of an IPv6 address are the global routing prefix, typically assigned by international registries and Internet Service Providers (ISPs). Bits 49 to 64 are the Subnet ID, responsible for subnet division, usually allocated by network administrators. The 6Diffusion algorithm effectively generates a greater number of Subnet IDs.

\begin{table}[H]
    \centering
\caption{COMPARISION OF CANDIDATE NEW PREFIX RATE AND GENERATION NEW PREFIX RATE UNDER THE /80 PREFIX}
         \scalebox{0.75}{
    \begin{tabular}{c>{\centering\arraybackslash}p{0.07\linewidth}>{\centering\arraybackslash}p{0.07\linewidth}>{\centering\arraybackslash}p{0.07\linewidth}>{\centering\arraybackslash}p{0.12\linewidth}>{\centering\arraybackslash}p{0.07\linewidth}>{\centering\arraybackslash}p{0.07\linewidth}>{\centering\arraybackslash}p{0.07\linewidth}>{\centering\arraybackslash}p{0.12\linewidth}}
         \multicolumn{9}{c}{/80 address prefix}\\
         \toprule
\textbf{Method}& \textbf{$N_{\text{c-pre}}$}& \textbf{$N_{\text{cn-pre}}$}& \textbf{$  r_{cn-pre}$}& \textbf{$N_{\text{cn-pre}}$ $per_{10k}$}& \textbf{$N_{\text{g-pre}}$}& \textbf{$N_{\text{gn-pre}}$}& \textbf{$  r_{gn-pre}$}& \textbf{$N_{\text{gn-pre}}$ $per_{10k}$}\\
         \midrule
            Entropy-IP\cite{foremski2016entropy} & 99789 & 99784 & 99.78\% & 9978.40 & 5130 & 5129 & 5.13\% & 512.90 \\
            \addlinespace
            6Tree\cite{liu20196tree} & 62131 & 19361 & 19.36\% & 1936.10 & 22403 & 12363 & 12.36\% & 1236.30 \\
            \addlinespace
            DET\cite{song2020towards} & 64476 & 24886 & 24.89\% & 2488.60 & 24600 & 14860 & 14.86\% & 1486.00 \\
            \addlinespace
            6Hit\cite{hou20216hit} & 12710 & 12355 & 9.03\% & 902.65 & 8095 & 8020 & 5.86\% & 585.94 \\
            \addlinespace
            6GCVAE\cite{cui20206gcvae} & 25731 & 25717 & 25.72\% & 2571.70 & 1845 & 1842 & 1.84\% & 184.20 \\
            \addlinespace
            6VecLM\cite{cui20216veclm} & 107149 & 60272 & 34.73\% & 3472.77 & 9655 & 2480 & 1.43\% & 142.89 \\
            \addlinespace
            6Forest\cite{yang20226forest} & 68318 & 66424 & 62.13\% & 6213.37 & 24683 & 24284 & 22.72\% & 2271.55 \\
            \addlinespace
            6GAN\cite{cui20216gan} & 196371 & 194525 & 93.44\% & 9343.54 & 37083 & 36567 & 17.56\% & 1756.41 \\
            \addlinespace
            AddrMiner\cite{song2022addrminer} & 70979 & 63945 & 64.16\% & 6416.32 & 37449 & 35541 & 35.66\% & 3566.23 \\
            \addlinespace
            HMap6\cite{hou2023search} & 75766 & 74241 & 80.71\% & 8070.55 & 39060 & 38948 & 42.34\% & 4233.94 \\
            \addlinespace
            6Scan\cite{hou20236scan} & 26062 & 25581 & 28.42\% & 2842.33 & 16825 & 16619 & 18.47\% & 1846.56 \\
            \bottomrule
            \addlinespace
            6Diffusion & 91136 & 90742 & \textbf{95.42\%} & \textbf{9542.45} & 43807 & 43636 & \textbf{45.89\%} & \textbf{4588.77} \\
        \bottomrule
    \end{tabular}
    
    }
\end{table}

\textbf{/80 address prefix.} As shown in Table V, under the /80 prefix, the 6Diffusion algorithm from this study generated the highest number of address prefixes in the candidate set, producing 9542 prefixes for every 10k candidate addresses. The 6Diffusion algorithm performed the best under the /80 address prefix, being able to generate the most address prefixes.

As illustrated in Table V, in terms of generating new active address prefixes under the /80 prefix, the 6Diffusion algorithm from this study again demonstrated the best performance, generating 4588 active address prefixes for every 10k candidate addresses.

The experiments compared the performance of 6Diffusion with other algorithms under /32, /48, /64, and /80 address prefixes, with particular attention to the candidate new prefix rate and the generation new prefix rate of each algorithm. The 6Diffusion algorithm performed more admirably in most scenarios. Although the 6GAN showed a slight edge in a few cases, the 6Diffusion was still able to generate new active address prefixes in a broader range of scenarios, thus offering a greater advantage in terms of address generation diversity. Compared to other algorithms, generative model algorithms demonstrated superior performance in diversity, further confirming the advantage of the 6Diffusion algorithm in generating a diverse range of addresses.

\section{CONCLUSION}

In this work, we explored the implementation of target generation algorithms to aid in address discovery within the global IPv6 address space. We proposed 6Diffusion, a generative network architecture based on diffusion models, designed to generate high-quality IPv6 candidate set addresses for scanning purposes.

6Diffusion utilizes diffusion model algorithms to proactively generate IPv6 addresses, enhancing the generation rate, activity rate, and the rate of generating new address prefixes. 6Diffusion semantically constructs the IPv6 address space and employs the GLF-MSA module to learn both global and local features of IPv6, better capturing the high-dimensional characteristics and distribution patterns of active IPv6 addresses, thereby generating active IPv6 addresses. 6Diffusion effectively addresses the low activity rate and concentration issues in IPv6 target address generation. Experimental results demonstrate that 6Diffusion outperforms state-of-the-art target generation algorithms across multiple metrics.

\section*{Acknowledgments}
The authors would like to greatly appreciate the anonymous reviewers for their insightful comments.

\bibliographystyle{IEEEtran}
\bibliography{references}

\end{document}